\documentclass[pdftex,twocolumn,10pt,letterpaper]{extarticle}

\newcommand{\AUTHORS}{Authors here test testing}
\newcommand{\TITLE}{Fast Userspace Networking for the Rest of Us}
\newcommand{\KEYWORDS}{Put your keywords here}
\newcommand{\CONFERENCE}{NSDI}
\newcommand{\PAGENUMBERS}{yes}       
\newcommand{\COLOR}{yes}
\newcommand{\showComments}{yes}
\newcommand{\comment}[1]{}
\newcommand{\onlyAbstract}{no}
\usepackage{tikz}
\usepackage{xcolor}
\newcommand*\circled[1]{\tikz[baseline=(char.base)]{
                \node[shape=circle,fill,inner sep=1pt] (char) {\textcolor{white}{#1}};}}
\newcommand*\circledsmall[1]{\tikz[baseline=(char.base)]{
                \node[shape=circle,fill,inner sep=1pt] (char) {\textcolor{white}{\footnotesize #1}};}}
\usepackage{soul}

\usepackage{ifthen}
\ifthenelse{\equal{\PAGENUMBERS}{yes}}{%
\usepackage[nohead,
            left=0.75in,right=0.75in,top=1in,
            footskip=0.5in,bottom=1in,     
            columnsep=0.33in
            ]{geometry}
}{%
\usepackage[noheadfoot,left=.75in,right=.75in,top=1in,
            footskip=0.5in,bottom=1in,
            columnsep=0.33in
	    ]{geometry}
}

\usepackage{textgreek}

\usepackage[most]{tcolorbox}

\usepackage{epigraph}
\usepackage[font={small,it},labelfont={small,bf},aboveskip=1pt]{caption}

\usepackage{pifont}
\usepackage{appendix}
\usepackage[compact]{titlesec}              
\titlespacing{\paragraph}{0pt}{*1}{*1}      
\titlespacing*{\section}{0pt}{1.5ex plus 1ex minus .2ex}{1.3ex plus .2ex}

\usepackage[inline]{enumitem}
\setlist{itemsep=0pt,parsep=0pt,nosep}             

\usepackage{fancyhdr}

\ifthenelse{\equal{\PAGENUMBERS}{yes}}{%
  \pagestyle{plain}
}{%
  \pagestyle{empty}
}

\usepackage[numbers,sort]{natbib}

\usepackage{booktabs}
\usepackage[binary-units=true]{siunitx}
\usepackage{multirow}
\usepackage{color}
\usepackage{colortbl}
\usepackage{float}                           

\usepackage[T1]{fontenc}
\usepackage[utf8]{inputenc}
\usepackage{libertine}
\usepackage[libertine]{newtxmath}
\usepackage{inconsolata}
\usepackage{textcomp}
\usepackage[mathcal]{euscript}

\newcommand{\mt}[1]{Machnet}

\definecolor{commentgreen}{RGB}{2,112,10}
\usepackage{listings}
\usepackage{mathtools}
\usepackage{algorithm}
\usepackage[noend]{algpseudocode}

\ifthenelse{\equal{\COLOR}{yes}}{%
  \PassOptionsToPackage{hyphens}{url}\usepackage[colorlinks,citecolor=blue]{hyperref}
}{%
  \usepackage[pdfborder={0 0 0}]{hyperref}
}
\usepackage{url}

\usepackage{cleveref}
\crefformat{section}{\S#2#1#3} 
\crefformat{subsection}{\S#2#1#3}
\crefformat{subsubsection}{\S#2#1#3}

\hypersetup{%
pdfauthor = {\AUTHORS},
pdftitle = {\TITLE},
pdfsubject = {\CONFERENCE},
pdfkeywords = {\KEYWORDS},
bookmarksopen = {true}
}

\usepackage[font={small,it}]{subfig}
\definecolor{placeholderbg}{rgb}{0.85,0.85,0.85}

\usepackage{tikz}
\usepackage{soul}
\usepackage{nicefrac}
\usepackage{setspace}

\newcommand{\ns}[1]{\SI{#1}{\nano\second}}
\newcommand{\us}[1]{\SI{#1}{\micro\second}}
\newcommand{\ms}[1]{\SI{#1}{\milli\second}}
\newcommand{\byte}[1]{\SI{#1}{\byte}}
\newcommand{\kbyte}[1]{\SI{#1}{\kilo\byte}}
\newcommand{\mbyte}[1]{\SI{#1}{\mega\byte}}
\newcommand{\gbyte}[1]{\SI{#1}{\giga\byte}}
\newcommand{\Gbps}[1]{\SI{#1}{Gbps}}

\newcommand{\rssminus}{\texttt{RSS-\hspace{0.1em}-}}

\usepackage[absolute]{textpos}

\newfloat{acmcr}{b}{acmcr}

\newcommand{\note}[2]{
    \ifthenelse{\equal{\showComments}{not}}{\textcolor{#1}{\small #2}}{}
}

\newcommand{\superscript}[1]{\ensuremath{^{\textrm{#1}}}}		

\usepackage{titling}
\newcommand*{\TitleFont}{%
      \usefont{\encodingdefault}{\rmdefault}{b}{n}%
      \fontsize{16}{20}%
      \selectfont}

\date{}
\title{\vspace{-40pt}\textbf{\TitleFont \TITLE}}
\author{Alireza Sanaee$^{1}$\textcolor{black}{\thanks{Equal contribution.}}, Vahab Jabrayilov$^{2}$\superscript{\textcolor{red}{*}}, Ilias Marinos$^3$\thanks{Work done while at Microsoft.}, Anuj Kalia$^4$\superscript{\textcolor{red}{\dag}},\\
Divyanshu Saxena$^5$, Prateesh Goyal$^6$, Kostis Kaffes$^2$, Gianni Antichi$^7$\\
\small{$^1$University of Cambridge, $^2$Columbia University, $^3$NVIDIA, $^4$OpenAI}\\ 
\small{$^5$The University of Texas at Austin, $^6$Microsoft Research, $^7$Politecnico di Milano}}

\usepackage{microtype}  

\begin{document}

\maketitle
\thispagestyle{empty}


\begin{abstract}
After a decade of research in userspace network stacks, why do new solutions remain inaccessible to most developers?
We argue that this is because they ignored (1) the hardware constraints of public cloud NICs (vNICs) and (2) the flexibility required by applications.
Concerning the former, state-of-the-art proposals rely on specific NIC features (e.g., flow steering, deep buffers) that are not broadly available in vNICs.
As for the latter, most of these stacks enforce a restrictive execution model that does not align well with cloud application requirements.

We propose a new userspace network stack, \mt{}, built for public cloud VMs.
Central to \mt{} is a new ``Least Common Denominator'' model, a conceptual NIC with a minimal feature set supported by all kernel-bypass vNICs.
The challenge is to build a new solution with performance comparable to existing stacks while relying only on basic features (e.g., no flow steering, no RSS reconfiguration).
\mt{} uses a microkernel design to provide higher flexibility in application execution compared to a library OS design; we show that microkernels' inter-process communication overhead is negligible on large cloud networks.
Our experiments show that \mt{} works on today's three largest public clouds.
We also demonstrate the latency and throughput benefits of \mt{} for two real-world applications: a key-value store and state-machine replication.
For the key-value store application, \mt{} achieves 80\% lower latency and 75\% lower CPU utilization compared to the best-existing cloud solution.


\end{abstract}

\ifthenelse{\equal{\onlyAbstract}{no}}{%
\section{Introduction}
\label{sec:intro}

Distributed applications running in data centers (e.g., distributed data stores, stream processing engines, caching systems, backends of productivity software) are hungry for fast inter-server communication~\cite{PRISM, Chardonnay2023OSDI, farm, kv-driect}.
When the developers of these applications look into kernel-bypass ("userspace") networking, a well-studied approach to reduce intra-machine networking latency, they often surprisingly realize that existing systems fail to meet their needs.

While userspace networking has been successful in large real-world deployments (e.g., with userspace TCP at Alibaba~\cite{luna}, Remote Direct Memory Access in Microsoft~\cite{Bai:nsdi23}, and Pony Express RPCs at Google~\cite{snap}), the successes have been limited to specialized internal services.
These systems require privileged access to specific hardware features and careful integration with target applications, such as hypervisors, remote storage, and bare-metal applications.
However, they are not generally compatible with the most common deployment model—virtual machines (VMs) in public clouds.
Even internal workloads at large companies like Amazon and Microsoft often run as guests on the public cloud (with a few exceptions) and, thus, cannot use these networking services~\cite{microsoft-on-azure}.
Even though many userspace networking systems (called "stacks" in this paper) have been made publicly available, such as  mTCP~\cite{mtcp}, libvma~\cite{libvma}, eRPC~\cite{erpc}, TAS~\cite{tas}, and Demikernel~\cite{demi-kernel}, they have not seen widespread adoption.
What discourages application developers from using one of the many open-source userspace stacks?

\begin{table*}[!t]
\centering
\resizebox{\textwidth}{!}{%
\small
\begin{tabular}{llll}
\textbf{NIC feature} & \textbf{Description} & \textbf{Approximate year of introduction} & \textbf{Network stacks that used the feature} \\
\toprule
Flow steering & Redirect matching packets to a specific receive queue & 2009 (Intel 82599ES) & eRPC~\cite{erpc}, Snap~\cite{snap}, R2P2~\cite{kogias2019r2p2} \\
RSS reconfiguration & Query/modify the RSS key or redirection table & 2009 (Intel 82599ES) & TAS~\cite{tas}, RSS++~\cite{rssplusplus}, mTCP~\cite{mtcp} \\
\midrule
TX DMA from app memory & Transmit packets directly from DMA-registered user memory & 2009 (Mellanox ConnextX2) & Cornflakes~\cite{cornflakes}, eRPC~\cite{erpc} \\
Remote DMA & Access remote server memory without involving remote CPU & 2009 (Mellanox ConnextX2) & SocksDirect~\cite{socksdirect}, mRPC~\cite{mrpc} \\
\midrule
Deep RX queues & RX queues with thousands of entries, to prevent host-side packet drops & 2009 (Mellanox ConnectX2) & eRPC~\cite{erpc}) \\
Multi-packet RQs & Receive multiple packets with one RX queue descriptor & 2014 (Mellanox ConnectX4) & eRPC~\cite{erpc}, Virtuoso~\cite{virtuoso}  \\
\bottomrule
\end{tabular}
}
\caption{LCD model has not changed in the past 15 years, long slack between hardware support and Cloud adoption. Examples of NIC features that are \textit{not exposed} to MVs created in public clouds of major providers. Existing solutions heavily rely on them since some of these features have been available for more than a \textit{decade} to this date.}
\label{table:unsupported_features}
\vspace{-0.1in}
\end{table*}

First, we find that virtual network interfaces (vNICs) used by cloud VMs, the standard environment for deploying modern applications, lack many features these stacks need.
Their restricted feature set stems from the public clouds' need to support (a) network virtualization and (b) a consistent interface across a decade of deployed NIC generations.
Notably, we find that no vNIC offered by leading cloud providers supports "flow steering," crucial for mapping network flows to CPU cores, a standard feature in bare-metal interfaces since 2009~\cite{intel-82599}. 
Other missing functionalities in vNICs include configurable support for receive-side scaling (RSS), RDMA, and multi-packet receive queues~(Table~\ref{table:unsupported_features}).
Although newer NIC models may support all these features, the financial impracticality of replacing older, widely deployed models prevents their universal availability in the near future~\cite{nic-purchase}.

Second, we identify that the architectural complexity of userspace stacks is also a limiting factor.
These stacks are based on either the library operating system (libOS)~\cite{erpc, mtcp} or sidecar~\cite{tas, snap} model.
The libOS model integrates the network stack directly into applications, leading to challenges such as the inability to enforce congestion control policies~\cite{shinjuku} and the need for applications to be written in low-level systems languages~\cite{belay2014ix, demi-kernel}.
The sidecar model addresses some of these issues by running the stack as a separate process, allowing greater flexibility.
Existing sidecars, though, co-design the components like CPU scheduler, queue management, and networking stack to squeeze the last drops of performance out of CPUs~\cite{shenango, snap, caladan}.
The complexities of these models and their extreme performance objectives—processing tens of millions of packets per second—are not only overkill for typical public cloud environments~\cite{twitter-trace} but also increase the deployment challenges for users who are not cloud infrastructure experts.
For instance, these bespoke systems often optimize using knowledge of application service time~\cite{demi-kernel, shenango, kogias2019r2p2}.
It is challenging to apply such configuration in a cloud environment by non-expert developers, and misconfiguration may result in wasted computing resources.

To work on the VMs across the major cloud providers and to enable easy adoption by non-expert cloud users, we have created a novel userspace network stack called \mt{}, built explicitly for cloud VMs.
We started by modeling a common vNIC to understand the features \mt{} can rely upon.
We created the Least Common Denominator (LCD) NIC model, the minimal feature set supported by all kernel-bypass vNICs.
Developers of userspace stacks who want to target broad applicability should use this model.
Our experiments show that this includes only plain Ethernet packet I/O, opaque RSS (e.g., without key inspection or modification), only as many NIC TX/RX queues as the number of CPU cores, and at most 256 descriptors per NIC queue.
To allow flexibility in application execution, \mt{} uses a simplified sidecar design where the stack process mediates access to the NIC from applications.
Our results show that due to the large size of cloud networks, the inter-process communication overhead of sidecars adds only 10\% to the p99 latency compared to a less flexible library OS design~(\S\ref{subsec:microbenchm}).

Our work's key challenge is to architect a performant stack while adhering to the constraints of the LCD NIC model.
For example, to scale \mt{} over multiple CPU cores, we design a new flow-to-core mapping technique called \rssminus{} that requires only the LCD NIC's opaque RSS functionality.
The academic community has long ignored the limited LCD feature set as even recent kernel bypass systems explicitly targeting cloud deployment, such as Junction~\cite{junction:nsdi24}, depend on non-LCD features.
These systems, built for cloud providers rather than users, require direct access to the NIC and, as a result, are unusable by application developers in the public cloud today.
We show that \mt{} performs well on the three major public clouds and evaluate its performance on two production-level applications: a key-value store and state-machine replication. \mt{} is currently open-sourced\footnote{https://github.com/microsoft/machnet}, and we will release the link if accepted.

The main contributions of this paper are:
\begin{itemize}
\item We make the case for a new userspace network stack specifically built for cloud VMs.
\item We build \mt{}, the first low-latency userspace network stack designed for cloud VMs and applications. \mt{} (1) relies on only the common set of NIC features available in all major clouds and (2) supports a flexible execution model that allows multiple processes and many threads sharing the same NIC, interrupt-based execution, and bindings for high-level languages.
\item We demonstrate the system in the wild on public clouds and evaluate a production-level key-value store and state-machine replication application. 
\end{itemize}

\section{Background and Motivation}
\label{sec:background}
We next discuss the main factor that has discouraged developers from using existing fast userspace stacks in public cloud VMs: incompatibility with cloud vNICs (\S\ref{sec:limitations-of-vnics}) and the needs of cloud applications (\S\ref{subsec:cloud-apps}).
We then detail the features an ideal portable stack should provide alongside characteristics we do not target (\S\ref{sec:requirements}).

\subsection{Incompatibility with cloud NICs}
\label{sec:limitations-of-vnics}
Cloud NICs (or vNICs) are a level of abstraction above bare-metal NICs and their virtual functions (VFs).
This is the fundamental reason why userspace stacks designed for bare-metal NICs and VFs might not work on vNICs.

An important factor in public clouds is that the fleet's network hardware spans multiple generations of hardware from multiple vendors, e.g., Microsoft's Azure cloud uses NICs spanning 12 years, from ConnectX-3 devices released in 2011~\cite{connectx3} to Microsoft's MANA devices released in 2023~\cite{mana, mana-kernel}.
To provide a consistent virtualized network with manageable complexity, the vNICs exposed to tenants typically provide a uniform feature set~\cite{gpoll, ec2poll, netvsc}.
Note that exceptions to this exist for specialized VM types targeting particular use cases (e.g., vNICs in VMs for High-Performance Computing support RDMA).

\begin{figure}[t!]
\centering
\includegraphics[width=0.9\columnwidth]{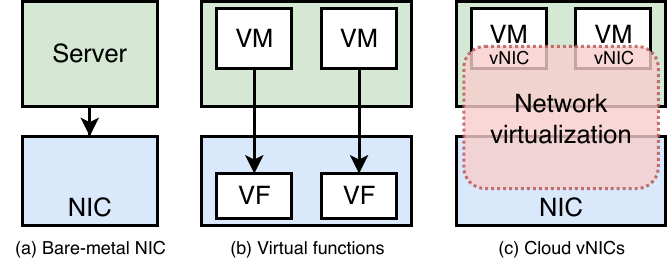}
\caption{A logical comparison of bare-metal NICs, virtual functions, and cloud vNICs.}
\label{fig:diff-nic}
\vspace{-0.1in}
\end{figure}

Figure~\ref{fig:diff-nic} shows a qualitative comparison of three types of bare-metal NIC, virtual functions (VFs), and vNICs.
Bare-metal NICs expose the NIC's entire suite of features to the user, ranging from low-level knobs like RSS and flow steering to entire protocol layers like RDMA and Transport Layer Security (TLS)~\cite{Pismenny:asplos21}.
VFs are isolated NIC slices that share the physical NIC's hardware resources~\cite{vf}, which typically expose most but not all the physical NIC's features~\cite{gpoll, ec2poll, netvsc}.
Simple virtualization environments (e.g., on-premise data centers) typically dedicate a VF to each VM; the VM can access this VF from the guest's userspace if needed.

In contrast, guest VMs in public clouds do not have raw access to the virtual function since that would bypass the cloud's network virtualization layer.
The abstraction layer provides network virtualization~\cite{Koponen:nsdi14,azure_smartnic,andromeda}, which includes (1) tenant features like bring-your-own-IPs, firewalls, routing, and access control lists, and (2) cloud operator features like inter-tenant isolation, transparent network upgrades, servicing, monitoring, and VM migration.

Since Microsoft has published details about their Azure vNIC implementation~\cite{vfp, azure_smartnic}, we use it as a representative example.
In their ConnectX-based vNICs, the network virtualization layer spans an FPGA datapath, host hypervisor software, and the guest VM's paravirtual device driver~\cite{azure_smartnic}.
During network I/O, guest applications must use Azure's paravirtual ``netvsc'' drivers for both kernel- and kernel-bypass I/O instead of the virtual function's corresponding raw ConnectX drivers.
The paravirtual driver includes support for network virtualization events such as servicing (e.g., for the FGPA SmartNIC), monitoring, VM migration, etc.
In AWS, their purpose-built Nitro devices implement this layer~\cite{nitro}.
Google Cloud Platform (GCP) uses dedicated CPU cores for network virtualization~\cite{snap}.

Existing fast userpace stacks often aim to exploit advanced features available in bare-metal NICs: this is the case for example of flow steering~\cite{mtcp, erpc,demi-kernel}, multi-packet receive queues~\cite{erpc,virtuoso}, or deep receive queues~\cite{erpc}, to name a few.
One may expect those stacks to work as-is on vNICs with minor modifications. Unfortunately, we find that this is not the case and significant design and implementation changes are required.
The problem is that while the features they rely upon have been ubiquitous in bare-metal NICs for over a decade (e.g., Intel's 82599 Ethernet controllers released in 2009 support flow steering~\cite{intel-82599}), they are not yet part of the virtualized NICs exposed to guest VMs.

In Table~\ref{table:unsupported_features}, we list the features used in existing projects that are not supported by a large span of NICs deployed currently in the cloud.

\vspace{0.1in}
\noindent\fbox{\begin{minipage}{23.5em}
\noindent{\textbf{Consequence.}} A userspace network stack for the cloud needs to leverage \textit{only} the features available at cloud vNICs. Those depend on the network virtualization layer that exposes only a small subset of the ones available from bare-metal NICs.
\end{minipage}}

\subsection{Incompatibility with cloud applications}
\label{subsec:cloud-apps}
Cloud applications display a range of characteristics that match the dynamic and distributed nature of modern computing environments~\cite{cortez2017resource}.
These applications are typically written in various programming languages, including high-level ones such as JavaScript, C\#, Go, and Python~\cite{scio-net, meta-microservices}.
They are logically composed of diverse components, including databases~\cite{Chardonnay2023OSDI}, key-value stores~\cite{memcached, memcached-meta}, web servers~\cite{CaddyGitHub}, and authentication services~\cite{HashiCorpVault}, each serving distinct functions.
Multiple components are often co-located in the same VM to minimize the communication cost~\cite{ibench-interference}.
Furthermore, administrators need to leverage oversubscription, effectively running multiple processes or threads per core to maximize resource utilization~\cite{cortez2017resource}.
As we more extensively discuss in~\S\ref{subsec:sidecar}, the problem is that most of the userspace networking stacks~(full system solutions such as Shinjuku, or Demikernel~\cite{shinjuku, demi-kernel}) follow the libOS model~\cite{erpc, cornflakes, mtcp}, effectively compiling the networking stack together with the application, which in turn takes full ownership of the NIC.
This model is incompatible with the characteristics of cloud applications.

\vspace{0.1in}
\noindent\fbox{\begin{minipage}{23.5em}
    \noindent{\textbf{Consequence.}} A userspace network stack for the cloud should not be integrated with applications to ensure support for multiple programming languages and the ability to serve multiple applications at the same time.

\end{minipage}}

\subsection{Requirements from a cloud-native userspace stack.}
\label{sec:requirements}

In the following, we summarize the characteristics an ideal userspace stack designed for the cloud should provide:
\begin{itemize}
    \item \textbf{R1. NIC agnostic.} The stack should not depend on any NIC features other than those available in the vast majority of cloud vNICs (\S\ref{sec:lcd-cloudvnic}).
    \item \textbf{R2: Support for diverse applications.} The stack should support many processes at the same time since rarely a VM hosts just a single application (\S\ref{subsec:sidecar}).
    \item \textbf{R3: Low-latency.} The stack should not compromise on performance: it should be comparable with existing approaches that rely on specific NIC features (\S\ref{subsec:low-latency}).
\end{itemize}


\paragraph{Non-goal 1: Highly communication-intensive workloads.}
\label{sec:non_roadblocks}
This paper does not target massive-scale communication-intensive applications like storage and machine learning, already served by RDMA and other fabrics~\cite{Bai:nsdi23, azurehpc, efa}.
Instead, we target the long tail of cloud applications like distributed databases, caches, and stream-processing engines.
From our discussion with developers, we have found that the key metric of interest in these applications is \emph{latency}.
For example, while existing userspace stacks projects target tens of millions of requests per second~\cite{erpc,demi-kernel,farm,kv-driect} (at the cost of restricted application execution models), production workloads are often far less demanding in terms of message rate: an analysis of Twitter's production cache workload shows that most servers run under 100k requests per second~\cite{twitter-trace}.
Google's Snap paper mentions a server handling a few million requests per second at peak load~\cite{snap}.

\ul{\textit{Implication.}} We are willing to sacrifice message rate and bandwidth in favor of compatibility with a wide range of vNICs and application execution models as long as latency remains competitive~\cite{tail-at-scale, killer, Chardonnay2023OSDI}.

\paragraph{Non-goal 2: Compatibility with unmodified applications.}
In our discussions with engineering teams, we found that few expect to run unmodified applications on a userspace stack.
Instead, developers are often willing to rewrite the networking parts of their application and, in some cases, even co-design the application with the stack.
This is intuitive for two reasons.
First, reducing latency often necessitates changes to the application's threading model, e.g., to avoid context switches, which go hand-in-hand with changes to the networking model.
Second, most applications do not directly use low-level networking APIs (e.g., POSIX) but project-specific middleware (e.g., gRPC or a hand-written messaging layer), limiting the rewriting effort to a small part of the code.

\ul{\textit{Implication.}} We do not target binary compatibility with the BSD sockets API.
An API that resembles it is sufficient.
\section{Design Principles}
\label{sec:design-principles}

In this section, we discuss the principles that drove our design. Those allow us to meet the requirements we set for a new cloud-native userspace stack (\S\ref{sec:requirements}).

\subsection{LCD model for cloud NICs}
\label{sec:lcd-cloudvnic}
We must understand how to model a cloud NIC to meet the first requirement.
We define the ``Least Common Denominator'' (LCD) model, the common set of features available in all modern kernel-bypass capable NICs, including the vNICs of the three major cloud providers (Amazon EC2, Microsoft Azure, and Google Cloud Platform).

Our insight is that for broad adoption of a stack, it must run seamlessly across bare-metal NICs, VFs, and vNICs; the LCD model captures the common features that stack designers can rely on.
The LCD NIC model (Table~\ref{tab:lcd-model}) provides OS-bypass Ethernet packet I/O over a number of TX and RX queues up to the number of CPU cores, with queues containing up to 256 descriptors each. Receive side scaling is supported, but this NIC does not expose the RSS key or RSS indirection table to the guest VM for inspection or modification. 

\begin{table}
    \centering
    \begin{minipage}{.9\columnwidth}
        \centering
        \fbox{
            \small
            \begin{tabular}{>{\bfseries}ll}
                I/O interface & Ethernet frames \\
                \#RX/TX queues & = number of CPU cores \\
                TX/RX queue size & 256 descriptors \\
                List of offloads & Opaque receive side scaling
            \end{tabular}
        }
    \end{minipage}
    \caption{Least common denominator NIC model in \mt{}}
    \label{tab:lcd-model}
\end{table}

Note that individual vNICs may support additional features beyond the LCD model and may be worth individually optimizing.
For example, most vNICs in Azure VMs support registering application memory with the NIC for zero-copy transmission (see below).
However, vNICs in other clouds do not support this feature (e.g., Google Cloud Platform, Amazon EC2), so we currently do not include it in the LCD model.

\paragraph{Queues and descriptors.}
Since a cloud provider must virtualize the NIC's queues and SRAM among tenants, each tenant gets a limited piece of these resources.
Across all three major cloud providers, we observed that the number of TX/RX queue pairs equals the number of server vCPUs rented by the tenant, and the number of descriptors per queue is 256.
This precludes designs like eRPC~\cite{erpc} that create very large receive queues to avoid host-side packet drops.
The LCD NIC requires posting and polling one descriptor per packet; descriptor coalescing optimizations such as batching multiple packets per descriptor~\cite{erpc} are unavailable.

\ul{\textit{Implication.}} 
The new userspace stack should work as a separate process, multiplexing the available NIC queues among multiple threads.

\paragraph{Flow-to-CPU core mapping.}
The LCD NIC supports only opaque receive side scaling (RSS), i.e., the NIC randomly hashes flows to RX queues, but the stack may not query or modify the NIC's RSS key or indirection table.
In contrast, granular control over bare-metal NICs' flow-to-core mapping units has been widely used to scale network stack throughput over multiple cores and to implement rich CPU scheduling policies.
For example, Snap~\cite{snap} and eRPC~\cite{erpc} install NIC flow rules that match on packet headers and redirect matched packets to a specific RX queue, which are then processed by a particular CPU core.
TAS~\cite{tas} and RSS++~\cite{rssplusplus} reconfigure the NIC's RSS indirection table to dynamically scale the number of CPU cores.
mTCP~\cite{mtcp} installs a special symmetric RSS key on the NIC~\cite{tas}.
IX~\cite{belay2014ix} queries the NIC's RSS key to choose UDP source ports for flows that maximize CPU affinity.

\ul{\textit{Implication.}} The new userspace stack should use only opaque RSS while providing isolation between applications (no performance inference is created).

\paragraph{Direct memory access.}
During packet I/O with the LCD NIC, the stack copies packets between application memory and the NIC's TX/RX ring buffers.
DMA-related optimizations are not available.
For example, in Cornflakes~\cite{cornflakes}, the NIC reads DMA-registered application memory to serialize objects into Protocol Buffers.
The benefits of RDMA have been widely studied in numerous systems~\cite{socksdirect, mrpc, farm}, but support for RDMA in guest VMs (e.g., for performance isolation~\cite{Kong:nsdi23}) is still in progress.

\ul{\textit{Implication.}} The new userspace stack should be built atop plain packet I/O instead of relying on zero-copy packet transmission or RDMA.

\subsection{Diverse application execution models}
\label{subsec:sidecar}
Userspace stacks have historically sacrificed application execution flexibility for performance.
For example, the libOS-based approach---used by Sandstorm, libvma, Seastar, OpenOnload, eRPC, and Demikernel~\cite{Marinos:sigcomm14, libvma, seastar, openonload, erpc, demi-kernel} ---has been the standard, with a few notable exceptions~\cite{tas, snap}.
The libOS approach imposes several constraints on the application's execution model, the most important of which is that one application ``owns'' the NIC, so only one process can use the NIC at a time.
Additionally, each application thread typically owns one NIC queue, so the small number of NIC queues on cloud vNICs limits the number of threads.
Note that simply attaching many vNICs to the cloud VM is not a feasible solution to these problems since it goes against the standard VM deployment model.
In Microsoft Azure, among general-purpose, computer-optimized, memory-optimized, and storage-optimized VMs, the maximum vNICs allowed for VMs is 8~\cite{azure-vnic}.
Further, Oracle cloud limits the number of vNICs to the number of vCPUs in a virtual machine~\cite{oracle-vnic}, and VMware only allows up to 10 vNICs regardless~\cite{vmware-vnic}.


For Internet service applications, like databases, caches, and stream-processing engines, the libOS model is too restrictive.
One server often runs multiple applications written in various languages, with possibly numerous threads per process.
For example, Google production workloads are incompatible with this model since they run hundreds of threads per core and use high-level languages such as Go~\cite{google-workload}.


\paragraph{No threading constraints.}
\label{par:no-threading-constraints}
There is a need to support an arbitrary number of threads to accommodate modern multi-threaded cloud applications.
This problem is challenging in libOS approaches that require applications to use a single-threaded event loop.
Particularly, the number of threads would be limited to either the number of NIC queues or the number of CPU cores available to the virtual machine.

\ul{\textit{Implication.}} The new userspace stack should adopt a sidecar (microkernel) approach. Here, the sidecar owns the NIC. With this layer of indirection, multiple processes and threads can use the same NIC.
The sidecar can send interrupts to applications permitting non-poll mode processing.

\paragraph{Blocking receive.}
\label{par:blocking-receive}
Userspace stacks typically require applications to poll for incoming messages continuously.
Although this approach gets the best latency, it restricts application execution flexibility by limiting the number of threads to the number of CPU cores.
The problem is that in production workloads, often many threads share just one core~\cite{google-workload}.
For example, one workload we learned about needed low latency when hosting thousands of database instances isolated in different processes on a single VM; most of these expected little traffic.
Each database needs low-latency messaging to replicate data to remote servers when active.
This execution model is ill-served by a polling model since the number of database processes exceeds the number of CPU cores.

\ul{\textit{Implication.}} The new userspace stack should allow users to block on receive calls, i.e., application threads can sleep and be woken up by the sidecar when a message arrives.

\subsection{Low-latency support}
\label{subsec:low-latency}
Linux, the most common OS in the cloud~\cite{azure_smartnic}, is notoriously slow, and existing research attempted to improve its latency with disruptive changes~\cite{terabit-ethernet}.
While with this paper, we attempt to tackle this problem, designing the new stack following a sidecar model (as noted before) might raise performance concerns about the overhead imposed by the inter-process shared-memory communication (IPC) when data has to be delivered to the application.
It is worth noting that the IPC is small, considering our target deployments in large cloud networks.
Indeed, as we show later, an extra \us{1} is significant in small bare-metal testbeds, e.g., it is 43\% of eRPC's \us{2.3} median latency, but cloud networks are large, with VMs often separated by multiple switches and long fiber-optic cables~\cite{Guo:sigcomm16}.
The extra \us{1} is relatively small, e.g., only 10\% higher than eRPC in our measurement (\S\ref{sec:evaluation}).

\section{\mt{}}
\label{sec:overview}
\begin{figure}[t!]
    \centering
    \includegraphics[width=0.80\columnwidth]{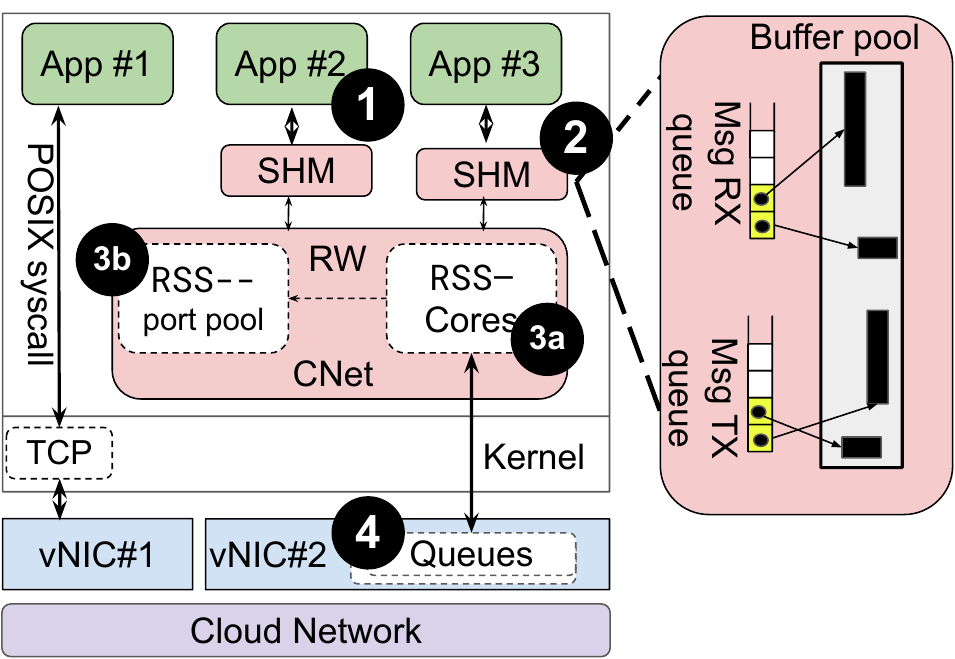}
    \caption{\mt{} architecture. Legacy application (App 1) can use traditional POSIX syscalls to communicate over Linux Kernel TCP while
    multiple applications~(App 2 and App 3) connect network over CNet using the shim library}
    \label{fig:machnet-overview}
    \vspace{-0.1in}
\end{figure}

\mt{} has two components: a ``sidecar'' userspace process that handles all kernel-bypass network I/O and a shim library that applications use to communicate with the sidecar (Figure~\ref{fig:machnet-overview}).
The shim library provides sockets-like API calls for applications to send and receive messages and communicates with the sidecar over shared memory channels, summarized in Appendix Table~\ref{tab:machnet_api_brief}.
The \mt{} sidecar process handles all kernel-bypass network I/O and can scale to multiple cores without relying on flow steering.

Applications initiate a shared memory channel with the sidecar process~\circled{1}, facilitating message exchange with the \mt{} sidecar.
From a security perspective, these shared memory regions are entirely isolated, and a compromised application cannot manipulate the shared memory region of other applications. 
Each shared memory channel~(SHM)~\circled{2} comprises a pair of Message RX/TX queues, managing references to the head of incoming and outgoing buffers.
\mt{} employs \rssminus{} powered "engines"~\circledsmall{3a} per CPU core to handle the processing of ingress/egress packets from associated RX/TX queues~\circled{4}.
During connection setup, \mt{} utilizes the \rssminus{} port pool~\circledsmall{3b} to associate the flow with the correct engine.
This association process occurs off the hot path (see \S\ref{subsec:rssminus} for further details).



\subsection{Scaling to multiple cores}
\label{subsec:multicore}
The \mt{} sidecar process can use multiple CPU cores to (1) scale performance to large VMs (e.g., with \Gbps{100} networking) and (2) provide performance isolation among applications.
We run one \mt{} ``engine'' thread per core; each engine has an exclusive NIC RX/TX queue pair and runs in a busy-polling run-to-completion loop.
We use a shared-nothing architecture for scalability (i.e., engines do not share flow state or shared memory channels).
One dedicated shared memory channel exists between each application thread and a \mt{} engine.
\mt{}'s shim library API also allows each application thread to specify the engine to use (the default policy is round-robin).
This is useful when binding a latency-critical application to a dedicated engine, while less latency-sensitive applications can share an engine.

\paragraph{Need for flow-to-core mapping.}
NIC-offloaded flow-to-core mapping is necessary to support (1) our shared-nothing multi-core architecture and (2) support for binding applications to specific engines.
Without this, a packet may land at any engine core.
This, in turn, requires shuffling packets or flows between engine cores, breaking the shared-nothing architecture and inter-application isolation.
For example, an ACK packet may land on a different engine core than the one that sent the original data packet, requiring the receiver to either shuffle the packet to the correct core or pull the flow state from that core.

Snap~\cite{snap} uses the NIC's flow steering hardware to map flows to engine cores.
TAS~\cite{tas} uses only RSS, but it cannot assign flows to specific engine cores, which is necessary for performance isolation between applications or flows.
In addition, TAS requires each engine to communicate with all application threads, which does not scale well to the large number of threads we target.
Regardless, the problem is that cloud VMs do not expose flow steering to guest VMs at the time of writing.

We found that all major cloud providers support RSS, a stateless NIC mechanism that hashes packet headers (e.g., a Toeplitz hash of the four-tuple) to a receive queue.
\footnote{Since we use one RX queue per core, we use core and RX queue interchangeably.}
RSS guarantees flow affinity to different cores, i.e., packets of the same flow will all hit the same receive queue.
But it is challenging to know {\it a priori} which queue a packet will land on, as discussed next.


\paragraph{Difficulty of inverting RSS.}
Can we use RSS for scaling \mt{} with its shared-nothing architecture?
We initially implemented a novel mechanism that requires knowing the RSS key since we initially believed retrieving the RSS key would be part of the LCD NIC model (e.g., DPDK provides an API to retrieve the RSS key).
While this approach worked on our initial set of test cloud VMs, we found that it is unreliable in practice.
Some cloud NICs refuse to expose the RSS key.
Others vary in the endianness of the computed RSS hash, introducing additional complexity by requiring hardware-specific code and testing paths.

Our mechanism worked as follows.
Below, we use a client-server connection as an example, where the server is listening on a particular UDP port.
During connection setup by the client, the server (i.e., the server-side \mt{} process) communicates its RSS key and number of RX queues to the client.
The client uses this and its RSS key to compute a UDP source port to hash all flow packets to the desired local and remote RX queues.
IX uses a local variant of this technique, where the client chooses a UDP source port that causes server-to-client packets to hash to the core that sent the original client-to-server packet~\cite{belay2014ix}.
However, the local variant alone cannot support the application-to-engine affinity we require since it cannot control which server-side RX queue the packet lands on.

\subsection{\rssminus{}}
\label{subsec:rssminus}
We created a new approach called \rssminus{} that uses randomization to work despite the opaqueness of RSS in cloud NICs.
\mt{} repeatedly tries to establish a connection using different source and destination ports until it finds a combination that hashes to the desired RX queues.
There are two exciting aspects of this approach.

\begin{figure}
    \centering
    \includegraphics[width=.8\columnwidth]{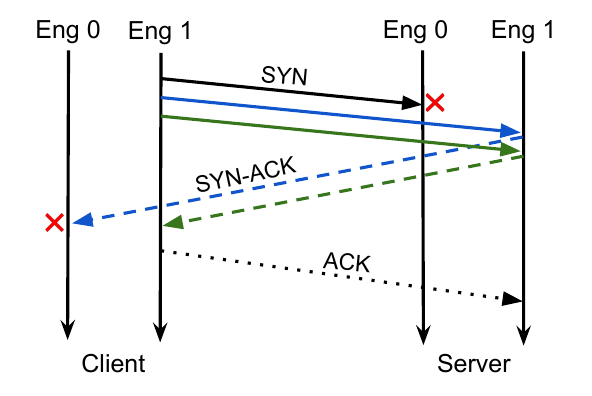}
    \caption{SYN and SYN-ACK spraying in \mt{} to establish a flow between engine \#1 at the client and engine \#1 at the server.
    This figure shows the unoptimized version where the client sprays $n^2$ SYN packets. The green SYN and SYN-ACK packets are the ones that hash to the correct RX queue.}
    \label{fig:machnet_handshake}
    \vspace{-0.1in}
\end{figure}

\paragraph{Decoupling flow identifiers from RSS.}
Unlike bare-metal NICs that allow offloading to the hardware advanced filters over arbitrary packet header fields, cloud NICs only support basic RSS hashing on the standard four-tuple for packet steering.
This couples RSS hashing and flow identifiers, limiting flexibility in how we can exploit randomization.
\mt{} breaks this coupling: instead of using UDP ports to define flows, we introduce two additional 16-bit port fields in the \mt{}-specific packet header.
As discussed below, this reduces the number of packets that \mt{} must spray during connection setup.
A side benefit is that it avoids reducing the number of supported flows between two servers as the number of RX queues increases.

\paragraph{SYN and SYN-ACK spraying.}
We first discuss an unoptimized method that requires a relatively large number of packets to establish a connection:
During the initial connection handshake, the client-side \mt{} sends a batch of SYN packets with randomly selected UDP source and destination port pairs (Figure~\ref{fig:machnet_handshake}).
The server responds with a corresponding SYN-ACK packet that hashes to the correct server RX queue for each SYN packet.
The client uses the first SYN-ACK packet received at the correct RX queue to establish the connection and discard the rest.
If no SYN packet reaches the correct engine, the client repeats after a timeout (\ms{300} by default) with different UDP port pairs.

How many SYN packets should the client send?
Let us assume two \mt{} machines with $n$ engines each and a uniformly random RSS hash function.
The probability of an SYN/SYN-ACK packet hashing to the correct server/client queue is $\frac{1}{n}$.
Thus, the probability of a connection being established by a given packet is $\frac{1}{n*n}$.
Using basic probability, the number of packets required to establish a connection with 95\% likelihood is $\log(1-0.95)/\log(1-\frac{1}{n*n})$.
This equals 47 packets for $n=4$ engines and 191  for $n=8$ engines.
The remaining 5\% of flows are established after a timeout and retry.

\paragraph{Reducing sprayed packets.}
With our decoupling of flow identifiers from RSS, the server does not simply need to reflect the client's SYN UDP port pair.
Instead, we can use a different UDP port pair for the reverse direction.
This does not change the \mt{}-specific port fields, so each side can still associate packets to the correct flow.
The client sends some SYN packets with random UDP port pairs in the optimized method.
On receiving the first SYN, the server responds with an equal number of SYN-ACK packets with its randomly chosen UDP port pairs.
As long as both the SYN and SYN-ACK packets hash to the correct remote RX queue with probability at least $\sqrt{0.95}$, we maintain the 95\% probability of connection setup.
This requires each side to send $\log(1-\sqrt{0.95})/\log(1-\frac{1}{n})$ packets; the total number of packets for connection setup is two times this value, which equals 25 and 55 packets for $n=4$ and $n=8$ engines, respectively.

\textit{Communicating successful UDP ports.}
The server and client communicate the successful UDP port pairs (used for future packets in the flow) to each other in the SYN-ACK and ACK packets, respectively.
The server includes the correctly-hashed SYN packet's UDP port pair in its SYN-ACK packets' payload.
Similarly, the client includes the successful SYN-ACK packet's UDP port pair in the subsequent ACK packet and first data packet.
Note that the client and server may use different UDP port pairs for each flow direction.

\section{Implementation}
\label{sec:implementation}
The \mt{} sidecar is implemented in C++ and uses DPDK for kernel bypass packet I/O.
We use C for the shim library to make interface with other languages easier.
The stack and shim library are implemented in \textasciitilde11K lines of code, excluding language bindings.

\paragraph{Why DPDK instead of XDP?}
\label{par:dpdk-xdp}

Interestingly, the same lack of flow steering that led us to develop \rssminus{} also cripples AF\_XDP in cloud VMs.
We discuss this next.
AF\_XDP is a recent Linux improvement that bypasses most kernel stack processing to reduce latency, which may be ``good enough'' for some applications.\footnote{
Another advance---io\_uring---focuses on improving throughput by batching syscalls, which is irrelevant to our latency target.}
In its best ``zero-copy'' case, the device driver receives a packet directly to user-space memory, bypassing the networking code.
We studied AF\_XDP in detail on cloud VMs and implemented a \mt{} variant based on DPDK's efficient AF\_XDP driver, and found deficiencies, inline with prior reports~\cite{karlsson2018path}.

\begin{figure}[t!]
    \centering
    \includegraphics[width=.8\columnwidth]{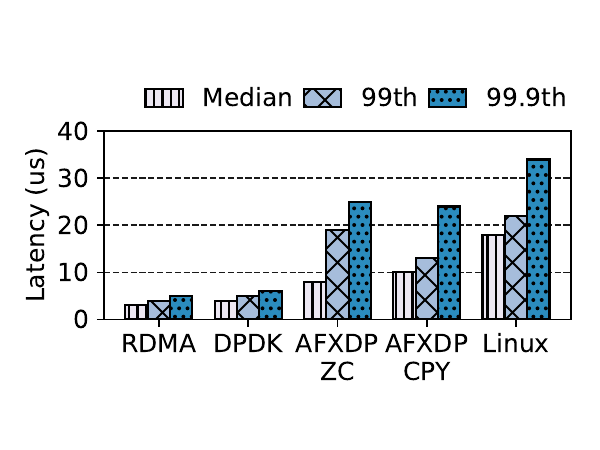}
    \vspace{-0.3in}
    \caption{This experiment measures the single message latency of each approach.
    DPDK performs significantly better than existing options.}
    \label{fig:afxdp-single-message-latency}
    \vspace{-0.1in}
\end{figure}

Surprisingly, we found that AF\_XDP does not use zero-copy in cloud VMs.
The reason is interesting: we found that \textit{\ul{zero-copy AF\_XDP requires flow-steering support in the NIC as a security measure}}, to prevent packets destined for one application from being delivered to another~\cite{Topel2018}.
Without zero-copy support, latency is not significantly better than that of the standard kernel network stack, and is up to far worse than with zero-copy.
Second, to support multiple application processes, AF\_XDP also requires a sidecar model (so it cannot avoid the IPC overhead) since the number of NIC queues limits the number of AF\_XDP sockets.

Zero-copy is the primary challenge when using AF\_XDP, but it is not the only factor contributing to performance issues. When an application uses AF\_XDP, the kernel is responsible for putting messages in the correct memory regions when received from the network and transmitting application messages placed in the shared memory region by the application. For the application to experience minimal latency, kernel threads must work hand in hand and in sync with the application threads to avoid extra latency. Unfortunately, this does not happen in practice, especially in VMs, where threads are not guaranteed to get the actual physical CPU from the hypervisor, and slight hiccups will cause high tail latency.

We ran experiments to understand the performance of AF\_XDP, using bare-metal nodes to allow experiments with zero-copy (xl170 Cloudlab).
We pin the interrupt (IRQ) and application threads on different CPU cores for best results.
The first experiment measures round-trip latency with a single message in flight using two nodes, comparing RDMA, DPDK, AF\_XDP zero-copy (ZC), AF\_XDP copy, and standard Linux.
Figure~\ref{fig:afxdp-single-message-latency} shows that DPDK achieves 4.1x and 5.6x lower latency at 99.9th percentile compared to AF\_XDP ZC and Linux.

\begin{figure}[t!]
    \centering
    \includegraphics[width=1\columnwidth]{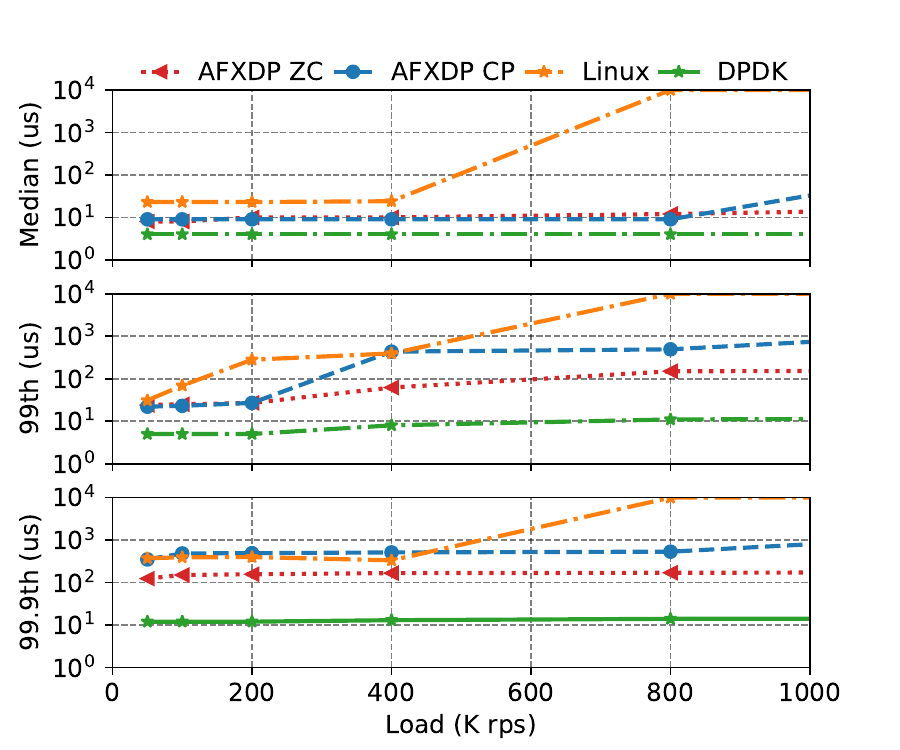}
    \caption{DPDK performs significantly better than AF\_XDP zero-copy.
    Unfortunately, even the best non-DPDK option (AF\_XDP ZC) is not available
    in cloud virtual machines due to the lack of flow steering support. }
    \label{fig:afxdp-load-latency}
    \vspace{-0.1in}
\end{figure}

The second experiment measures the load latency of each approach via three nodes.
A client on one node only measures the latency by sending single messages, and another workload generator residing on another node produces the background workload of 64-byte messages.
A separate server node echoes messages back to the clients.
We vary the background load and measure the latency.
Figure~\ref{fig:afxdp-load-latency} shows each approach's median, 99th, and 99.9th percentile latency.
DPDK dominates AF\_XDP by 10x at 99.9th percentile.

\paragraph{Network protocol.}
\label{par:network-protocol}

Cloud NICs favor connection-oriented protocols over connectionless protocols.
When a guest VM initiates a new connection, an SDN policy evaluation pipeline processes the control packets (e.g., SYN)~\cite{vfp}.
While this is transparent to the VM, it significantly increases first-packet latency and limits the number of connections per second that can be initiated~\cite{sirius, andromeda}.
It is common for packets-per-second to drop by 100$\times$ or more when initiating new connections~\cite{Bansal:nsdi23, vfp}.
Moreover, connectionless transports in combination with packet spraying via random UDP ports (e.g., Homa~\cite{homa}, NDP~\cite{NDP}) may run into SDN-related bottlenecks.
For instance, Homa uses short-lived connections to enable its spraying mechanism.
Unfortunately, every new connection created should go through an SDN slow path to fill up SDN flow tables, resulting in degraded performance down to 10K packets per second.~\cite{Bansal:nsdi23}.

To reduce the SDN overheads of the public cloud, we design \mt{} to feature a connection-oriented protocol similar to TCP.
\mt{} provides reliable in-order message delivery, with fragmentation and reassembly and selective acknowledgments for retransmissions.
We use UDP packets with an added \mt{} header, implementing a flow-based message-oriented protocol that preserves message boundaries (e.g., unlike TCP's byte stream).
We support message sizes up to \mbyte{8} and allow multiple outstanding messages per flow.
Overall, other transport protocols, as well as TCP, can be implemented in \mt{}.

\paragraph{Packaging.}
We package \mt{} as a Docker container so that users need not deal with DPDK, which is often difficult to install and run.
DPDK is a large project with around two million lines of code, an unstable API, and dynamically-probed drivers, making its complexity comparable to a small operating system.
As an example, a common usability problem occurs as follows: DPDK does not build the driver for ConnectX NICs if it does not find the NIC's userspace libraries; when this happens, applications build fine but fail with a cryptic error message at runtime when DPDK fails to locate a driver for the NIC.
To further complicate matters, incompatible versions of these libraries are available from three different sources (upstream rdma-core, the distribution's rdma-core, and Mellanox OFED), causing further confusion.

While these may seem minor issues, we believe this complexity has been a serious roadblock for adopting library OS approaches, which require application developers without DPDK backgrounds to go through a steep learning curve.
Our \mt{} container image includes a pre-built sidecar binary along with all required DPDK patches(e.g., Google Cloud Platform patch\cite{compute-virtual-ethernet-linux}), which can be downloaded and run with a small shell script.
Application developers using \mt{} do not interface with DPDK but instead use the sockets-like shim library API.

\section{Evaluation}
\label{sec:evaluation}

\begin{table}
\centering
\begin{tabular}{lllll}
\toprule
\textbf{Cloud provider} & \textbf{Size} & \textbf{p50} & \textbf{p99} & \textbf{p99.9} \\
\toprule
Microsoft Azure & \byte{64} & 27 & 32 & 49 \\
\cline{2-5}
    & \kbyte{32} & 81 & 97 & 159 \\
\midrule
Amazon EC2 & \byte{64}& 48 & 53 & 57 \\
\cline{2-5}
    & \kbyte{32} & 224 & 240 & 257 \\
\midrule
Google Cloud & \byte{64} & 65 & 111 & 164 \\
\cline{2-5}
    & \kbyte{32} & 221 & 273 & 335 \\
\bottomrule
\end{tabular}
\caption{\mt{}'s round trip latency (\textmu{}s) for echo messages. The values across cloud providers should not be directly compared.}
\label{tab:all-three}
\vspace{-0.1in}
\end{table}

We show that \mt{} meets all three requirements set in \S~\ref{sec:requirements}.
In \S~\ref{subsec:cloud-platforms}, we show that it is compatible with the vNICs of all three major cloud providers (AWS, Azure, GCP).
In \S~\ref{subsec:faster}, \ref{subsec:raft}, \ref{subsec:app-isolation}, we show that \mt{} is compatible with the diverse application execution models of cloud applications.
Finally, in \S~\ref{subsec:microbenchm}, we show that \mt{} achieves latency comparable to or better than other userspace network stacks that do not adhere to the LCD model.
Throughout the evaluation, we compare \mt{} to three alternatives, noting that \mt{} is a network stack-only solution. We compared it against existing network stacks rather than full operating systems approaches, including Shenango~\cite{shenango}, and Junction~\cite{junction:nsdi24}~(\S\ref{disc:fops}):

\vspace{0.1cm}
\noindent\textbf{Linux TCP/IP.} We use this as a representative of OS-based approaches.
We use Sockperf~\cite{sockperf} for our experiments.
Libraries built atop this (e.g., gRPC) will have higher latency and lower throughput than Linux TCP/IP.

\vspace{0.1cm}
\noindent\textbf{eRPC.}~\cite{erpc} eRPC is not an LCD-compatible network stack, and it only runs on a limited number of Azure VMs that support RSS reconfiguration.
We use this (commit b8df1bc) to represent library-based stacks (e.g., Demikernel).
Unlike \mt{}, eRPC supports only one process per NIC and only one thread RX queue (= CPU core).

\vspace{0.1cm}
\noindent\textbf{TAS.} We use this~\cite{tas} (commit d3926ba) as a representative for sidecar-based stacks (e.g., Snap, which is closed-source).\\
\ul{\textit{Differences.}} Unlike \mt{}, TAS's sidecar process uses at least two CPU cores, one for the fast path and one for the slow path.
This presents a common problem for small 1--8 core cloud VMs (e.g., Cortez et al. reported that more than 95\%  of Azure VMs are 8-core or smaller~\cite{cortez2017resource}).
While TAS supports multiple application processes, it does not provide performance isolation among them (\S~\ref{subsec:app-isolation}) since TAS lacks a technique like \rssminus{}.

\noindent \textbf{Why no Snap}. Unfortunately, it cannot be directly compared due to its proprietary nature. Also, Snap’s broader scope overlaps with \mt{} only in the PonyExpress component.

\subsection{Evaluation setup}
\label{subsec:mcloud}
To demonstrate our LCD NIC model fits all modern DPDK-capable Ethernet NICs, we have tested \mt{} on various cloud and bare-metal NICs.
\mt{} works on Azure, AWS, and GCP, and it has been tested on several bare-metal NICs, including ConnectX-3 to ConnectX-6 and Intel E810.

\paragraph{Configuration.} Unless mentioned otherwise, we evaluate \mt{} on a cluster of VMs on Azure's public cloud.
Most of our optimization effort went to Azure, which provides the best latency (\S\ref{subsec:eval-latency}).
The VMs are in the same region and availability zone in all experiments.
The VMs have accelerated networking and run Ubuntu 22.04.2 with Linux kernel 6.2.
We do not enable proximity placement groups to keep results generally applicable, which can further reduce latency between VMs in the same availability zone.
Our experiments use eight-core F8s\_v2 VMs with \gbyte{16} DRAM.
We chose small VMs since those represent the large majority of VMs in public clouds, e.g., \citet{cortez2017resource} report that over 98\% of Azure VMs have eight or fewer CPU cores.

Unless mentioned otherwise, we use the smallest configuration for the \mt{} and TAS sidecars, i.e., one engine core for \mt{} and two cores for TAS (one each for its fast and slow path).
For the Linux TCP/IP stack, Nagle's algorithm is disabled.
We use huge pages for the DPDK-based stacks (i.e., \mt{}, TAS, and eRPC) and pin stack processes to CPU cores.
We omit the more invasive tuning techniques, such as isolated CPU cores (\texttt{isolcpus}) or interrupt steering.

\subsection{Supported cloud platforms}
\label{subsec:cloud-platforms}

Table~\ref{tab:all-three} shows that \mt{} successfully runs on all three major public clouds, achieving our primary goal of creating a userspace stack for this environment.
We used AWS's ``ENA express''\cite{aws-ena} networking and Google Cloud's newer gVNIC DPDK driver\cite{google2023gvnic}.
For the experiment, we measure the round-trip latency of \byte{64} and \kbyte{32} echo messages between two VMs in the same availability zone, each running a single-threaded echo application.
\mt{} achieves excellent tail latency on Azure and EC2, with the 99.9th percentile latency below \us{57} for \byte{64} messages.

Note that the latencies across cloud providers should not be directly compared since we have not yet used comparable VM types or hardware generations.
For example, we enabled placement groups to reduce the inter-VM distance on Amazon EC2 but not on the other clouds.
We believe that large-scale evaluation of userspace networking on public clouds---now possible with \mt{}---is an important direction for future work.

\subsection{FASTER key-value store over \mt{}}
\label{subsec:faster}
\begin{figure}[t!]
    \centering
    \includegraphics[clip,scale=0.5]{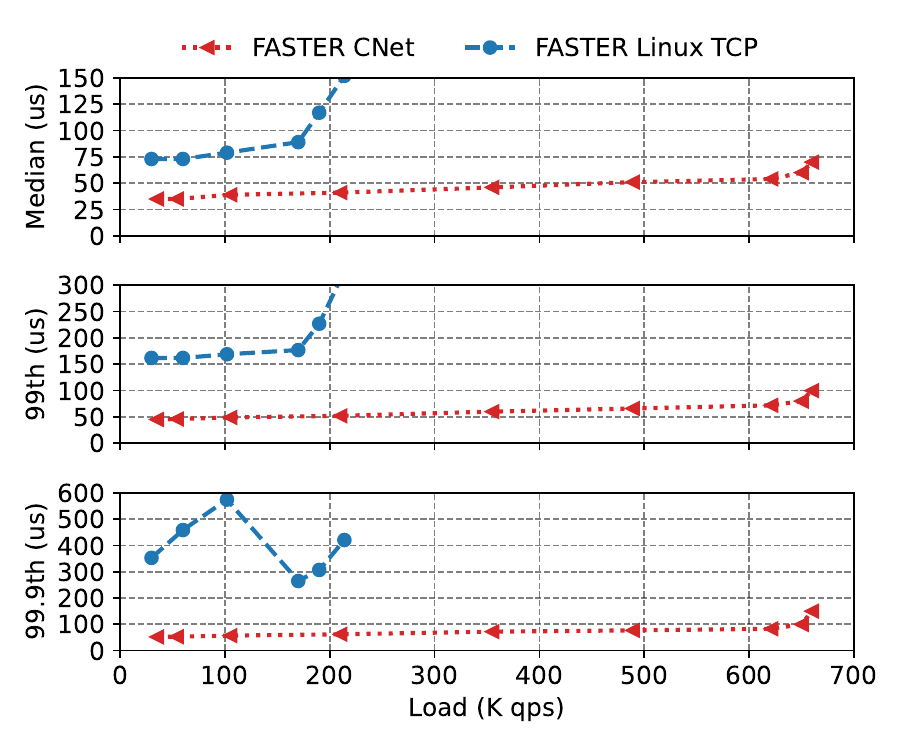}
    \caption{Performance of the FASTER key-value store over \mt{} and Linux TCP/IP.
    \mt{} achieves 3.3x higher throughput and 80\% lower p99 latency compared to Linux.}
    \label{fig:faster}
    \vspace{-0.1in}
\end{figure}

We next demonstrate \mt{}'s performance benefits and ease of use in end-to-end applications.
In all cases, we make no changes to the core application code and only modify their networking calls to use \mt{}.

FASTER is a production-grade high-performance key-value store~\cite{faster-kv,faster-github}.
For this experiment, we created one single-threaded server application that uses FASTER's C++-based in-memory hash table for storage (commit 0116754) and \mt{} for networking.
We pre-populate the server's hash table with 100 million eight-byte key-value pairs.
Two client VMs generate the workload, which consists of 100\% GET requests to emulate a read-heavy workload.
We vary the load by (1) increasing the number of concurrent connections from each client VM from one to 20 and (2) increasing the number of outstanding requests per connection from one to eight.
A third client VM acts as our latency probe, using one thread to send GET requests to the server one at a time.

\paragraph{Single-threaded performance.}
Figure~\ref{fig:faster} compares the throughput and latency achieved over \mt{} and Linux's TCP/IP stack.
The \mt{}-based implementation achieves 3.3x higher throughput, with 700K RPS compared to Linux TCP's 210K RPS.
At 210K RPS, \mt{}'s p99 latency is 80\% lower, achieving \us{50} compared to Linux TCP's \us{250}.
Another crucial improvement is in CPU utilization.
At its peak throughput, the in-kernel TCP stack causes 100\%  CPU utilization on all eight cores.
In contrast, the \mt{}-based implementation uses 100\% of only two CPU cores (application and one engine), i.e., 75\% lower.

\paragraph{Multi-threaded performance.}
\label{subsec:eval-multi-core}

\begin{figure}
    \centering
    \includegraphics[width=0.7\columnwidth]{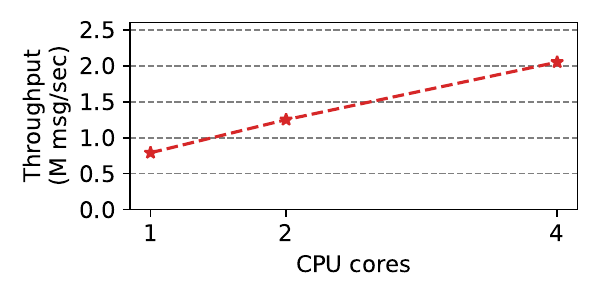}
    \caption{FASTER server performance with multiple threads}
    \label{tput:par:mmps}
    \vspace{-0.5cm}
\end{figure}

We evaluate \mt{}'s scalability to multiple engine threads based on \rssminus{} (Section~\ref{subsec:rssminus}) as follows.
The FASTER server application uses four threads, each pinned to a different CPU core.
We then run separate experiments with between one and four \mt{} engine threads.
We use four client VMs to generate the workload, to ensure no queuing at client side.
Figure~\ref{tput:par:mmps} shows that \mt{} scales well with the number of engines, handling 0.8 million requests per second (Mrps) with one engine and 2 Mrps with four engines.

\subsection{Golang-based Raft over \mt{}}
\label{subsec:raft}

State machine replication (SMR)~\cite{raft, paxos-made-simple,vr} is an essential part of cloud applications, used in various applications ranging from configuration management control planes~\cite{zookeeper,etcd,aws_tiny_dbs,delos}, to data-intensive systems~\cite{spanner,dynamodb,cockroachdb} that manage petabytes of data and millions of requests per second.
In SMR, each server in a cluster of servers runs an identical sequence of commands, allowing the service to withstand failures.
Since SMR protocols are often on the critical path of data-intensive systems, they must deliver low latency.
Production-grade implementations of SMR are highly complex, so integrating them with \mt{} is a good test of \mt{}'s generality.

For our study, we chose Hashicorp's production Raft implementation~\cite{hashicorp_raft} for two reasons.
First, it is written in a non-systems language (Go), which helps us demonstrate \mt{}'s broad applicability.
Second, it provides a generic network interface that permits different implementations.

\paragraph{Golang microbenchmark.}
We use cgo to implement Golang bindings for \mt{}'s shim library.
To evaluate its performance, we use a simple client-server echo application written in Go and compare its performance with its C++ counterpart for latency and throughput.
We use a similar setup as in Section~\ref{subsec:microbenchm}.
Our experiments show that the two perform similarly, with Golang performing slightly worse due to cgo's memory copy overhead.
For \kbyte{32} echo messages, the Go version has only 8\% higher p99 latency and 5\% lower throughput.

\paragraph{Raft benchmark.}
We compared our port of Hashicorp's Raft implementation to \mt{} with Hashicorp's official Linux TCP implementation.
Our Raft setup uses one client that issues commands (\byte{80} each, similar to eRPC's evaluation~\cite{erpc}) to the Raft cluster, which replicates them to an in-memory log.
We use three VMs, two running the SMR protocol (one leader and one follower) and one serving as a client load generator.
The client sends a request to the leader and waits for a response; the leader responds after replicating the data to its follower.
Each server VM has two active CPU cores: one for the application and one for \mt{}.

\begin{figure}[t!]
    \centering
    \includegraphics[clip,scale=0.5]{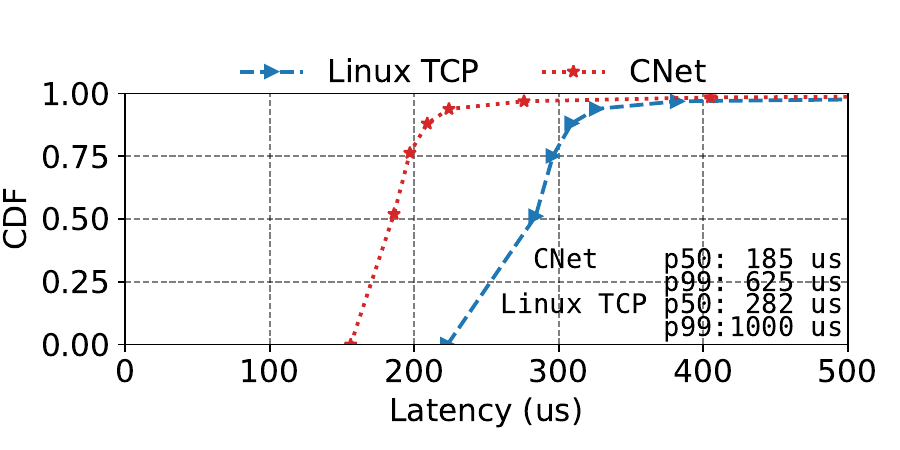}
    \caption{Client-measured latency for Hashicorp's Golang-based Raft}
    \label{fig:raft}
    \vspace{-0.5cm}
\end{figure}

Figure~\ref{fig:raft} shows that \mt{} outperforms Linux kernel TCP, with 34\% lower median latency and 37\% lower median p99 latency.
With \mt{}, we achieve microsecond-scale latency even at the 99th percentile, where the Linux kernel TCP counterpart exceeds a millisecond.

\subsection{Supporting diverse execution models}
\label{subsec:app-isolation}

\textbf{Performance isolation} \mt{} avoids inter-application performance interference by using \rssminus{} to map flows to specific \mt{} engines.
TAS supports only random flow-to-engine mapping (determined by the NIC's randomized RSS) and, thus, suffers from interference.
Our experiment to demonstrate this works is as follows.
We run four single-threaded echo applications at a server VM: three throughput-intensive and one latency-sensitive.
Client VM\#1 generates high-rate traffic to the throughput-intensive applications, using 16 flows with eight \byte{64} in-flight messages per flow.
Client VM\#2 acts as a latency probe, sending one \byte{64} echo at a time to the latency-sensitive application.
Since only the sidecar-based approaches support multiple concurrent applications, we exclude eRPC from this evaluation.

\begin{figure}
    \centering
    \includegraphics[clip,scale=0.8]{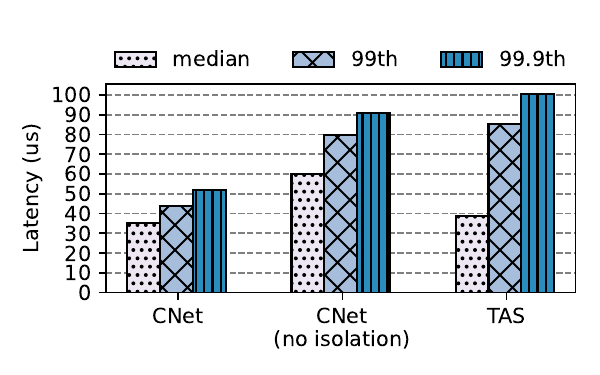}
    \caption{Inter-application performance isolation in \mt{}}
    \label{fig:eval-iso}
    \vspace{-0.1in}
\end{figure}

We use two engines for \mt{} at the server, one paired to the three throughput-intensive applications and one dedicated to the latency-sensitive application.
For TAS, we use two fast-path cores to match \mt{}'s fast-path cores in addition to TAS's one slow-path core.
Figure~\ref{fig:eval-iso} shows that with \mt{}, client VM\#2 achieves nearly 50\% better tail latency than TAS.
We also show the latency with \mt{} without isolation, i.e., when flows are randomly mapped to engines.
Without isolation, \mt{}'s tail latency, too, suffers from interference, showing the importance of \rssminus{}.

\textbf{Blocking receive.} We have found that a typical application pattern is one where there are more application threads needing network I/O than the number of CPU cores (Section~\ref{par:no-threading-constraints}), which requires support for blocking receive.
We evaluate \mt{}'s blocking receive with an experiment where a \mt{}-based echo server application runs multiple (up to four) threads, all sharing a single CPU core.
A single-threaded client application on a different VM generates a workload with \byte{64} request-response messages.

To measure latency, we configure the open-loop client to send load at full rate to the server.
Figure~\ref{fig:block-latency} shows that when the server applications use blocking receive, the client gets low tail latency even when the application threads share a single core.
In contrast, with multiple non-blocking server threads polling at the same CPU core, tail latency spikes to over \ms{10}.
This happens because a server thread may need to wait multiple Linux scheduling intervals before running.

Figure~\ref{fig:cpu-util} shows that blocking receives reduces CPU utilization.
We run the server with four threads for this experiment and configure the client to generate requests in an open loop, varying between 70K--700K requests per second (RPS).
With blocking receives, the server core's CPU utilization stays below 75\% even when handling 700K RPS.
In contrast, the CPU utilization of busy-polling is constantly 100\%, regardless of the load.
\subsection{Microbenchmarks}
\label{subsec:microbenchm}


\begin{figure}[t!]
\centering
\includegraphics[clip,scale=0.6]{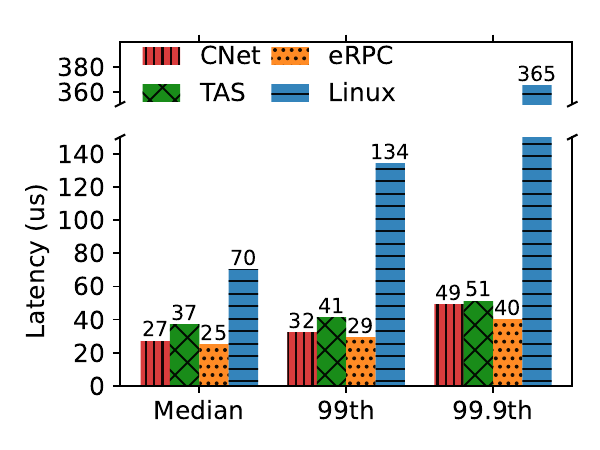}
\caption{Round-trip latency for 64-byte messages.}
\label{fig:eval-micro-lat}
\vspace{-0.3cm}
\end{figure}

\label{subsec:eval-latency}
\textbf{Latency microbenchmark.} Figure~\ref{fig:eval-micro-lat} compares \mt{}'s latency with the alternatives in the latency experiment above, with \byte{64} messages.
We make two observations.

First, \mt{}'s latency is comparable to eRPC, despite \mt{}'s use of only the LCD NIC model and support for diverse execution models.
Specifically, \mt{}'s median and p99 latency is within 10\% of eRPC and 23\% at p99.9.
Taking median latency as the example, \mt{}'s \us{27} latency is \us{2} higher than eRPC\footnote{Demikernel~\cite{demi-kernel} reports \us{15} lower round-trip latency on Azure. Our latencies are different because we use a different region.}.
This extra latency comes primarily from SHM communication between the application and \mt{}'s sidecar process, which costs around \ns{250} for each one-way SHM crossing.
Interestingly, while this overhead is significant compared to a small bare-metal testbed (e.g., eRPC reports \us{2.3} within a rack), in large cloud data centers, the network's base latency dwarfs this overhead.

Second, Linux TCP's latency is 2.6x, 4.2x, and 7.4x higher than \mt{} at median, p99, and p99.9, respectively.
This confirms prior findings that kernel-bypass's latency benefits, arising from avoiding context switches and heavy kernel processing, are significant even in cloud datacenters~\cite{demi-kernel,Chardonnay2023OSDI}.


\textbf{Throughput microbenchmark.} Figure~\ref{fig:eval-micro-tput} compares the throughput of the alternatives.
The client sends variable-sized messages to the server, which echoes them back.
The client maintains eight outstanding messages, so the throughput in this experiment also depends on latency.
\mt{} saturates the VM's available network bandwidth (\Gbps{12} each way) with \kbyte{8} messages.
Linux's throughput for message sizes up to \kbyte{8} is low because of its higher latency; for larger \mbyte{100} messages, Linux also saturates the network.

TAS's throughput scales well with message size, but it achieves ~20--40\% lower throughput than \mt{} for message sizes larger than \kbyte{2}.
eRPC, although highly optimized for small RPCs and small networks (e.g., sub-\us{10} RTT), requires further transport-related configuration tuning for better throughput with larger RPCs on cloud networks (e.g., for its per-flow credits).
In this experiment, we ensured that eRPC's flows were not credit-starved.

\begin{figure}[t!]
    \centering
    \hspace{-0.7cm}
    \includegraphics[clip,scale=0.8]{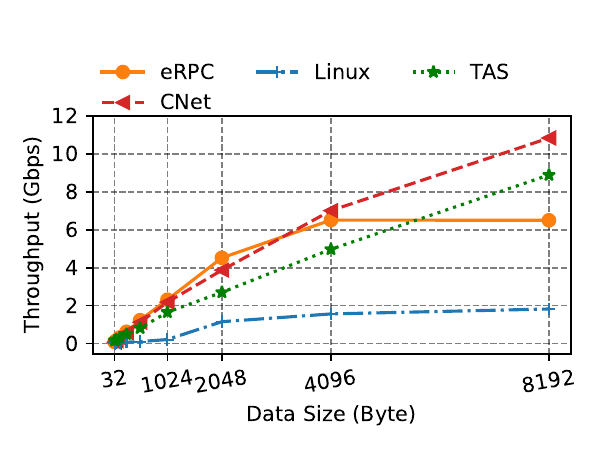}
    \caption{Unidirectional throughput for variable message sizes.}
    \label{fig:eval-micro-tput}
    \vspace{-0.1in}
\end{figure}

\textbf{\rssminus{} connection setup latency.} On top of the network round-trip time, \mt{}'s connection setup latency is largely dictated by the configurable periodic interval that \mt{} processes new connection requests by the applications; for our experiments, this is set to 50us.
Furthermore, when multiple engines are present, \mt{} relies on \rssminus{}, a packet spraying technique, to establish connections. If a connection handshake fails to be completed on the first batch of sprayed packets, \mt{} retries with a larger batch of packets to increase entropy. On such rare occasions, the retry timeout is inflating the connection setup latency.
We run a microbenchmark to present the connection setup latencies with \mt{}; we use bare-metal machines on both ends to avoid variations due to cloud-related network latency spikes.
We vary the number of engines at client and server ends from 1 to 8~(these numbers configure NIC queues corresponding to the engines, too) and measure the latency of connection setup at the application by creating 10K distinct connections one at a time.

\begin{figure}[t!]
    \centering
    \includegraphics[clip,scale=0.6]{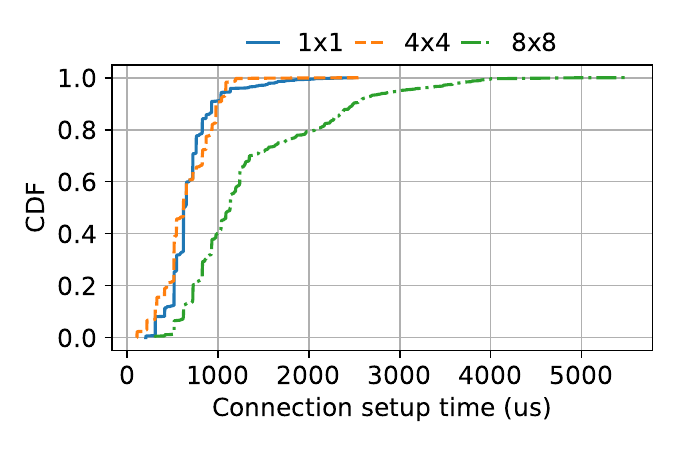}
    \caption{Connection setup latency in \mt{}. In all cases, connection setup time at the median is less than $\sim 1$ millisecond. (legend example: 4x4 means the sender machine uses four engines, and the receiver machine uses four engines).}
    \label{fig:machnet-conn-setup}
    \vspace{-0.1in}
\end{figure}

Figure~\ref{fig:machnet-conn-setup} shows the results. At the median, \mt{}'s \rssminus{} can achieve less than 1.5 millisecond connection setup latency.
The connection setup latency slightly increases, particularly at the tail, as the likelihood of selecting the wrong SRC port number is higher.
If no SYN packet from the initial batch reaches the remote \mt{} engine, or if none of the response SYN-ACK packets reach the \mt{} sender engine, there will be a SYN timeout and retry with a new exponentially larger batch of initial SYN packets.


\begin{figure}[t!]
    \centering
    \subfloat[Tail latency]
    {
        \hspace{-0.9cm}
        \centering
        \includegraphics[clip,width=0.26\textwidth]{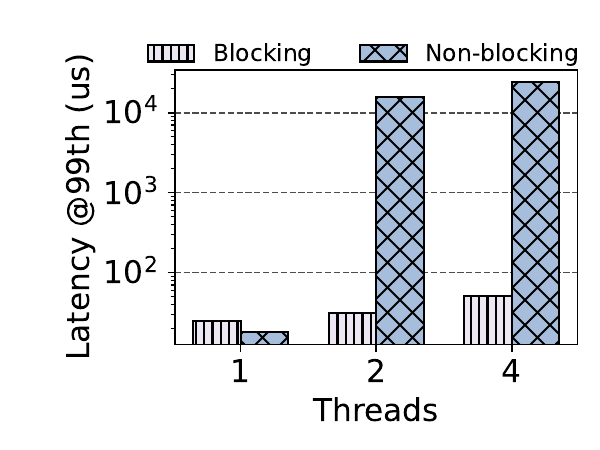}
        \label{fig:block-latency}
    }
    \subfloat[CPU utilization]
    {
        \centering
        \includegraphics[clip,width=0.26\textwidth]{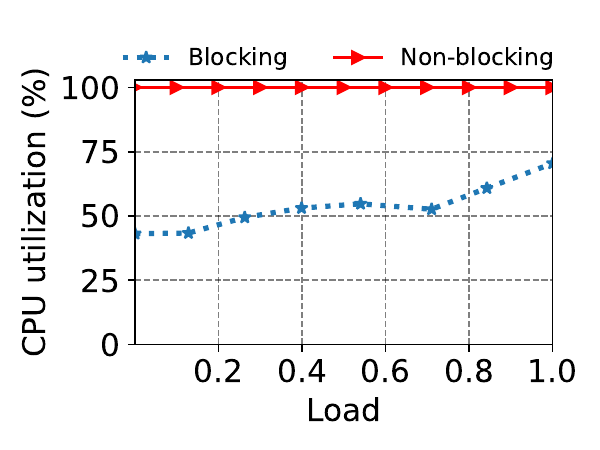}
        \label{fig:cpu-util}
    }
    \caption{Effectiveness of \mt{}'s blocking receives.}
    \label{fig:blocking-vs-non-blocking}
    \vspace{-0.2cm}
\end{figure}


\section{Discussion}
\label{sec:discussion}

\noindent \textbf{What happens if the LCD model changes?}
We derive the LCD model after analyzing the capabilities of the network virtualization layer exposed in the three major cloud providers.
It should be noted that large cloud providers typically spend millions of dollars to equip their fleet with a particular version of NICs~\cite{nic-purchase}. 

Table~\ref{table:unsupported_features} shows that the LCD model has seen little progress in 15 years, with features such as flow steering, available since 2009, still not accessible to cloud customers. Although the LCD model may improve as cloud providers upgrade NICs, we prioritized fast networking today, making RSS-- essential.


Nonetheless, \mt{} should adapt to the vNIC features available in its environment. \mt{} partial supports selective feature selection, using hardware checksum offloads (AWS and Azure) and software checksums in GCP where offloads are unavailable. Adding support for more features requires significant engineering, which we are addressing with the open-source community.



\vspace{0.5em}
\label{disc:fops}
\noindent \textbf{\mt{} and existing kernel bypass operating systems.}
A large body of research had focused on building full operating systems to support microsecond-scale applications in data centers such as Junction~\cite{junction:nsdi24}, ZygOS~\cite{zygos}, IX~\cite{belay2014ix}, Caladan~\cite{caladan}, Shenango~\cite{shenango}, and Shinjuku~\cite{shinjuku}. These systems are full-stack solutions, while \mt{}, being a networking stack, is an orthogonal effort. Indeed, \mt{} can be incorporated into these systems to provide more efficient transport and higher compatibility in the public cloud.

\vspace{0.5em}
\label{disc:custom_protocol}
\noindent \textbf{Does custom protocol hamper wider adoption?}
\mt{} targets cloud-based, latency-critical RPC workloads, using RDMA-like message-oriented transport suited for request-response tasks, avoiding TCP-like abstractions and overhead. Future plans include adding TCP for external service connectivity. Experiments used custom transport, with RSS-- orthogonal to transport protocols and implementable on stock TCP.
\section{Conclusion}

We created \mt{} after we got tired of answering ``no'' to application developers who asked ``Does userspace stack X work on the cloud?''
Initially, we believed this would be a simple matter of porting an existing stack to handle all the quirks of cloud vNICs.
However, it required a fundamentally new way of thinking about vNIC features as absent-by-default instead of present-by-default as in stacks designed for bare-metal NICs.
This paper attempts to crystallize our learnings---formed after numerous mistakes where we depended on a NIC feature that was unavailable in some vNIC on some cloud (e.g., RSS key inspection, application memory DMA registration)---into the Least Common Denominator NIC model.
Another crucial lesson was that the libOS model for userspace stacks is incompatible with how developers of non-infrastructure applications wish to run applications, i.e., with multiple processes and many threads, in various programming languages, and without touching DPDK.
We hope the research community will build upon \mt{}, with future projects like large-scale evaluations and new applications, to finally bring userspace networking benefits to the cloud.\label{lastpage}

\clearpage


\setlength{\bibsep}{2pt}
\bibliography{paper}
\bibliographystyle{abbrvnat}

\normalsize
\titleformat{\section}%
  {\bf}{Appendix~\thesection.\quad}{0pt}{}
\appendix
\appendix{Appendix}
\label{appendix}

\section{API calls and descriptions} APIs in \mt{} is devised to closely resemble traditional BSD-like socket APIs summarized at Table~\ref{tab:machnet_api_brief}.

\begin{table}[t!]
\centering
\caption{Summary of Machnet API Functions}
\resizebox{\columnwidth}{!}{%
\begin{tabular}{|l|l|}
\hline
\textbf{API Function} & \textbf{Description} \\ \hline
\texttt{machnet\_init()} & Initializes the library. \\ \hline
\texttt{machnet\_attach()} & Initializes the shared memory. \\ \hline
\texttt{machnet\_bind()} & == BSD bind() \\ \hline
\texttt{machnet\_listen()} & == BSD listen(). \\ \hline
\texttt{machnet\_connect()} & == BSD connect(). \\ \hline
\texttt{machnet\_send()} & == BSD send(). \\ \hline
\texttt{machnet\_recv()} & == BSD recv(). \\ \hline
\end{tabular}
}
\label{tab:machnet_api_brief}
\end{table}

\section{Future Opensource plans} We are planning to open source \mt{} in the near future. We are expecting wider adoption for this project since we are reducing the time to engage effort from deployment to development on this project.

}{
}

\end{document}